\documentclass[12pt]{iopart}

\usepackage{iopams} 

\usepackage{hyperref}
\usepackage{latexsym}
\usepackage{graphicx, amssymb, float,color}

\usepackage{appendix}

\usepackage{cite}

\newcommand{\calv}{C_{\mathrm{A}}}
\newcommand{\cart}{C_{\mathrm{a}}}
\newcommand{\cven}{C_{\bar{\mathrm{v}}}}
\newcommand{\cinh}{C_{\mathrm{I}}}
\newcommand{\cmeas}{C_{\mathrm{meas}}}
\newcommand{\cper}{C_{\mathrm{per}}}
\newcommand{\crpt}{C_{\mathrm{rpt}}}

\newcommand{\lrpt}{\lambda_{\mathrm{b:rpt}}}
\newcommand{\lper}{\lambda_{\mathrm{b:per}}}
\newcommand{\hen}{\lambda_{\mathrm{b:air}}}

\newcommand{\qper}{q_{\mathrm{per}}}
\newcommand{\qalv}{\dot{V}_{\mathrm{A}}}
\newcommand{\qc}{\dot{Q}_{\mathrm{c}}}

\newcommand{\prl}{k_\mathrm{pr}^{\mathrm{rpt}}}
\newcommand{\prm}{k_\mathrm{pr}^{\mathrm{per}}}
\newcommand{\ml}{k_\mathrm{met}^{\mathrm{rpt}}}
\newcommand{\mm}{k_\mathrm{met}^{\mathrm{per}}}

\newcommand{\valv}{\tilde{V}_{\mathrm{A}}}
\newcommand{\vrpt}{\tilde{V}_{\mathrm{rpt}}}
\newcommand{\vper}{\tilde{V}_{\mathrm{per}}}
\newcommand{\vinh}{\tilde{V}_{\mathrm{I}}}

\newcommand{\di}{\mathrm{d}}

\newcommand{\area}{\mathcal{A}}
\newcommand{\thickness}{\ell}
\newcommand{\solubility}{\beta}
\newcommand{\diffusivity}{\mathcal{D}}

\newcommand{\diffcap}{D}
\newcommand{\palv}{P_{A}}
\newcommand{\pblood}{P_{b}}

\begin{document}

\title[The role of mathematical modeling in VOC analysis]{The role of mathematical modeling in VOC analysis using isoprene as a prototypic example}

\author{H~Koc $^{1,3}$, 
        J~King $^{2,3}$,   
        G~Teschl $^3$, 
        K~Unterkofler $^{1,2}$,  
        S~Teschl $^4$,
        P~Mochalski $^{2,5}$,
        H~Hinterhuber $^6$ and
        A~Amann $^{2,7}$    }

\address{
$^1$Vorarlberg University of Applied Sciences,
Hochschulstr.~1,
A-6850 Dornbirn, Austria

$^2$ Breath Research Institute, Austrian Academy of Sciences,
Rathausplatz~4,
A-6850 Dornbirn, Austria

$^3$ Faculty of Mathematics, University of Vienna,
Nordbergstr.~15,
A-1090 Wien, Austria


$^4$ University of Applied Sciences Technikum Wien,
H\"ochst\"adtplatz~5,
A-1200 Wien, Austria

 $^5$ Institute of Nuclear Physics PAN, Radzikowskiego 152, PL-31342 
 Krak\'{o}w, Poland
 
 $^6$ Univ.-Clinic of Psychiatry, Innsbruck Medical University,
  Anichstr.~35,
 A-6020 Innsbruck, Austria

$^7$ Univ.-Clinic for Anesthesia, Innsbruck Medical University, 
Anichstr.~35,
 A-6020 Innsbruck, Austria
}

\ead{\mailto{anton.amann@i-med.ac.at}, \mailto{anton.amann@oeaw.ac.at}}

\begin{abstract} 
Isoprene is one of the most abundant endogenous volatile organic compounds (VOCs) contained in human breath and is considered to be a potentially useful biomarker for diagnostic and monitoring purposes. However, neither the exact biochemical origin of isoprene nor its physiological role are understood in sufficient depth, thus hindering the validation of breath isoprene tests in clinical routine. 

Exhaled isoprene concentrations are reported to change under different clinical and physiological conditions, especially in response to enhanced cardiovascular and respiratory activity. Investigating isoprene exhalation kinetics under dynamical exercise helps to gather the relevant experimental information for understanding the gas exchange phenomena associated with this important VOC.

A first model for isoprene in exhaled breath has been developed by our research group. In the present paper, we aim at giving a concise overview  of this model and describe its role in providing supportive evidence for a peripheral (extrahepatic) source of isoprene. In this sense, the results presented here may enable a new perspective on the biochemical processes governing isoprene formation in the human body. 

\end{abstract}

\ams{92C45, 92C35}

\submitto{Journal of Breath Research}

\noindent{\it Keywords\/}: Breath gas analysis, isoprene, volatile organic compounds, modeling
\maketitle

\section{Introduction}

\subsection{Why modeling is important}

In human breath more than 200 volatile organic compounds (VOCs) can be detected at trace level down to the part-per-trillion (ppt) range
 \cite{Pauling:1971, Phillips:1994, Amann:Book, Ligor:2008, Bajtarevic:2009, Herbig:2009, Kushch:2008a}. 
Endogenous VOCs are released within the human organism (e.g., as a result of normal metabolic activity or due to pathological disorders), enter the blood stream and are eventually metabolized or excreted via exhalation, skin emission, urine, etc..

Exhaled breath analysis has the advantage of being non-invasive. Breath samples can be extracted as often as desired and    can  be measured in real time, even in  breath-to-breath resolution \cite{Amann:2005, Amann:2009, King:isoprene, King:2009, King:PhysiolMeas, King2010a}.
This renders  breath analysis  as an optimal choice for  gaining continuous information on the current metabolic and physiological state of an individual \cite{Amann:2007, Miekisch:2004, Kharitonov:2002}.  
 
Identification and quantification of potential disease biomarkers can be seen as the driving force for the analysis of exhaled breath. Moreover, future applications for medical diagnosis and therapy control with dynamic assessments of normal physiological function or pharmacodynamics  are intended \cite{Amann:2004, Amorim:2007}.  
Exogeneous VOCs penetrating the body as a result of  environmental exposure can be used to quantify body burden   \cite{Pleil:2008}. 
Also breath tests are often based on the ingestion of  isotopically labeled precursors, producing isotopically labeled carbon dioxide and potentially many other metabolites \cite{Modak:2007, Modak:2009, Mattison:2004, Modak:2005, Modak:2010, Modak:2007a, Ezzeldin:2004}. 

Breath analysis is a young field of research and has not yet been widely accepted in clinical routine. Breath sampling, in particular,  is far from being a standardized procedure due to the numerous confounding factors biasing the concentrations of volatiles in breath \cite{Amann:2010}. These factors are related to both the breath sampling protocols as well as the complex physiological mechanisms underlying pulmonary gas exchange. Even under resting conditions exhaled breath concentrations of VOCs can strongly be influenced by specific medical parameters (such as cardiac output and breathing patterns), depending on the physico-chemical properties of the compound under study \cite{Cope:2004, Anderson:2006}. 

\begin{figure}[H]
\centering
\begin{tabular}{c}
\includegraphics[width=11.5cm]{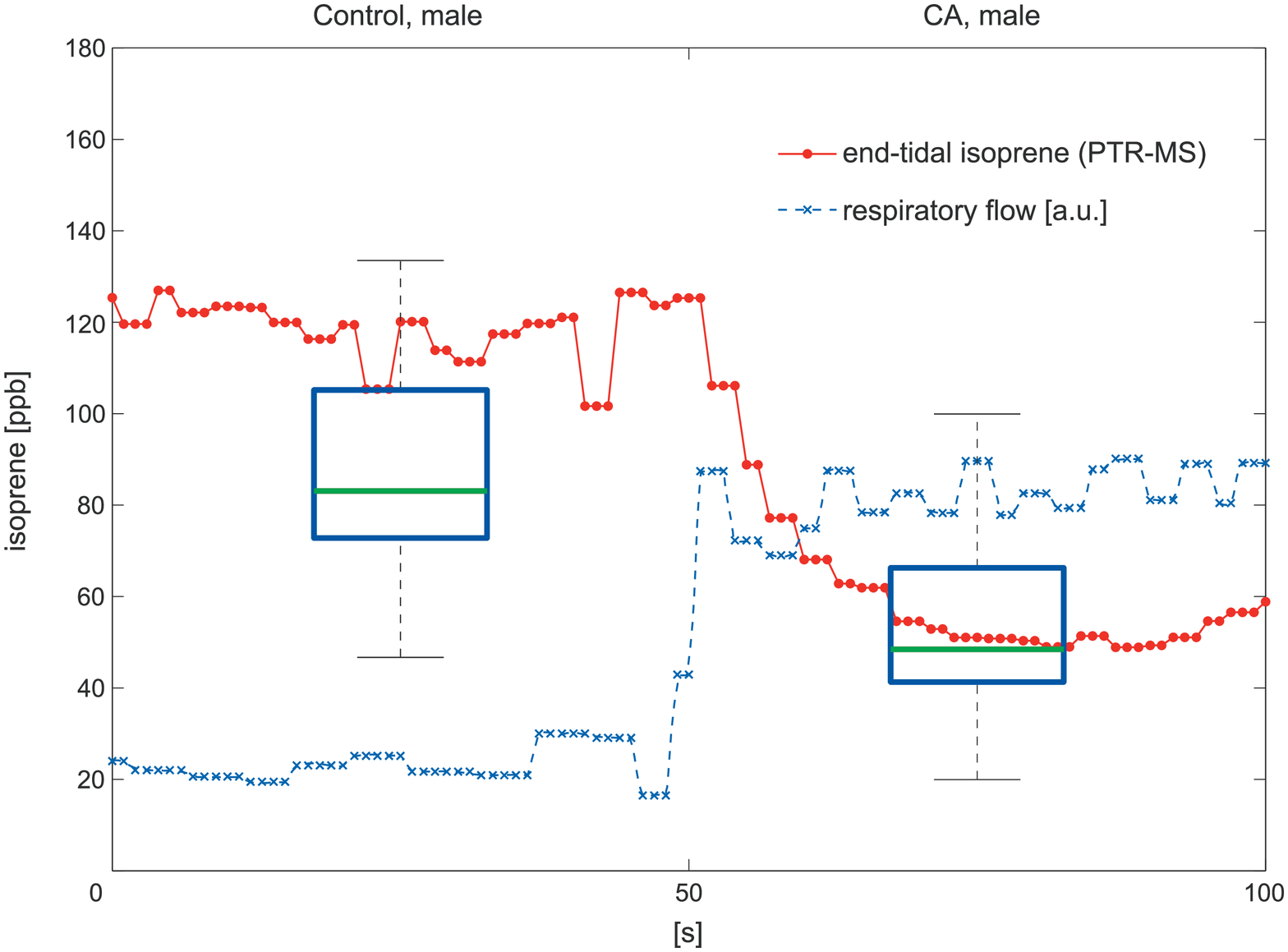}
\end{tabular}
\caption{Overlay depicting the variability of end-tidal isoprene concentrations during hyperventilation of one single volunteer as compared to the population boxplots associated with healthy test subjects and lung cancer patients, cf. \cite{Bajtarevic:2009}.}\label{fig:modeling}
\end{figure}

In order to obtain diagnostically conclusive results,  a thorough understanding of the physiological events and phenomena affecting exhaled VOC levels as well as an appropriate choice of breath sampling conditions and protocols are essential. 
For illustration of  this point, consider the population study result in Figure~\ref{fig:modeling}, which may suggest the detection of lung cancer patients on the basis of decreased isoprene levels in end-tidal breath.

Effectively, from the respective boxplots a tentative threshold value of about 70~ppb  differentiating between lung cancer patients and healthy volunteers might be established. It should be noted, however, that from a procedural point of view this observation will be of limited use, since any further attempt to classify a specific individual will strongly depend on the hemodynamic and respiratory status of the test subject investigated. Correspondingly, by measuring end-tidal isoprene levels with breath-to-breath resolution (e.g., using  proton transfer reaction-mass spectrometry (PTR--MS), cf.~\cite{King:2009}), it can be demonstrated that a normal healthy volunteer at rest might easily switch between the two groups defined above merely by changing his breathing pattern, e.g., by breathing faster or by increasing tidal volume. Doing so leads to an instantaneous drop of breath isoprene levels, thereby hampering an unambiguous classification of this volunteer. Such issues are further aggravated by the fact that human subjects tend to hyperventilate when they are asked to breathe normally and provide a breath sample~\cite{Risby:2008}. Consequently, any screening study results like the one above must be complemented by information on their variability with changes of ventilation, blood flow or pre-measurement conditions. 

This paradigmatic example shows that a proper quantitative assessment of the underlying exhalation kinetics is mandatory when aiming at the successful introduction of a clinically applicable breath test. 

In this context, mathematical modeling and simulation can be employed for capturing the decisive quantitative features of the observable data. The primary task of mathematical modeling hence is to provide a mechanistic description of the physiological phenomena governing the VOC under scrutiny using all the experimental evidence available. Once such a model has been developed it becomes possible to understand the key relationships underlying the physiological behavior  of the compound and to identify its defining parameters. Such a quantitative approach can aid substantially in preventing misinterpretations of experimental results and allows for the discovery of errors and omissions in earlier interpretations. Furthermore, these analyses can guide the standardization and establishment of new sampling protocols, avoiding potential confounding factors and maximizing the information content of experimental results.

\subsection{Compartmental modeling}\label{sec:comp}
Physiological models describe the behavior of substances by means of compartmental mass conservation. The compartments are defined as functional units of the organism, corresponding either to specific, localized anatomic structures (like liver, lung, kidney, etc.) or to types of tissues (such as fat, muscle, viscera, bone, etc.) that are distributed throughout the body \cite{Reddy:Book}. The concentration time course of a compound in each compartment is subsequently described by a mass-balance differential equation.

In the basic model describing  pulmonary gas exchange of blood-borne inert gases, the lung is assumed to be a single homogenous compartment with fixed effective  storage volume of the gas considered (see  \ref{sec:general}). For inert gases which are not present in inspired air and for steady state conditions, this leads to the well-known Farhi equation \cite{Farhi:1967a}

\begin{equation}\label{farhi}
\calv = \frac{\cven}{\hen + \qalv/\qc}\,. 
\end{equation}

It expresses the fact that the concentration of an inert gas in the alveolar air depends on the mixed venous concentration $\cven$, the substance-specific blood:air partition coefficient $\hen$, and the ventilation-perfusion ratio $\qalv/\qc$.
For the gas exchange between alveolar gas and capillary blood, it is generally postulated  
that the diffusive resistance across the alveolar-capillary membrane is the limiting factor, while limitations due to the diffusive conductance in alveolar gas are negligible \cite{Hlastala:Book}.
Although the relative equilibration of gas with blood along the pulmonary capillary depends on the balance between diffusing capacity ($\diffcap$, see \ref{sec:diff}) and perfusion,
for inert gases the exchange occurs very rapidly so that an instantaneous equilibration of end-capillary blood and alveolar air is regarded a reasonable premise in persons not suffering from a lung-related disease \cite{Hlastala:Book,Wagner:1974}.

 From the physical factors influencing gas exchange, solubility (i.e., the blood:air partition coefficient $\hen$) is the most important one because it determines the location of gas exchange.  
In classical theory, gas exchange is restricted to the alveolar space by reference to the assumption that no inert gas is added to or withdrawn from the alveolar gas as it flows through the conducting airways during inhalation and expiration. This is not true, however, for highly soluble gases, which may readily dissolve in the bronchial lining fluid and the blood of the bronchial circulation. Particularly, Anderson et.~al.~\cite{Anderson:2003} demonstrated that only
low soluble gases exchange almost exclusively in the alveoli, whereas  highly soluble gases exchange predominantly in the airways. As a result, diffusion of gas through the airway tissue, breathing patterns, bronchial blood flow, and inspired air conditions become the decisive factors determining overall pulmonary gas exchange. A corresponding model for the highly soluble compound acetone has recently been presented in  \cite{King2010a}.

In the present work we will restrict our attention to isoprene,
which is the most abundant hydrocarbon exhaled by humans. Although isoprene can be seen as a representative example for low-soluble endogenous VOCs, its exhalation kinetics show a characteristic behavior which is not yet understood in sufficient depth.  Conventional knowledge and available mathematical models describing the gas exchange mechanisms of the lung are
not able to clarify isoprene exchange mechanisms in a physiologically consistent manner.
A first model towards a new interpretation  
has recently been published by our research group \cite{King:isoprene}, based on the hypothesis of a peripheral source of endogenous isoprene formation. Here, we want to give an overview of the physiologically relevant facts driving model development and 
 present some insightful experiments carried out in an attempt to specify the location of the above-mentioned peripheral source.

\section{Experiments}
\begin{figure}[H]
\centering
\begin{tabular}{c}
\includegraphics[scale=0.68]{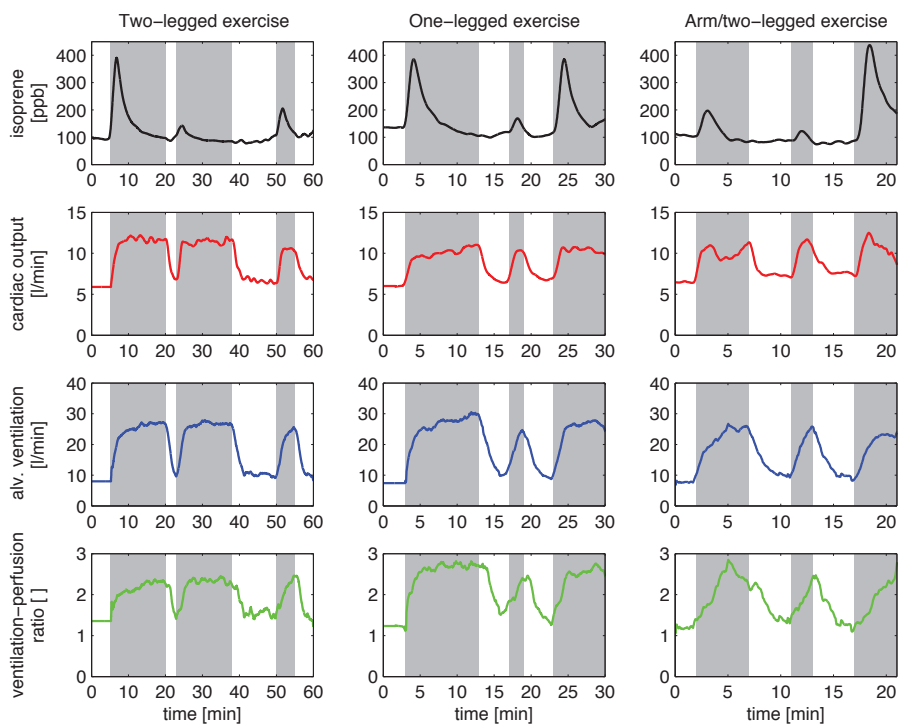}
\end{tabular}
\caption{Typical smoothed profiles of end-exhaled isoprene concentrations and physiological parameters in response to several workload  scenarios of one single volunteer (27 years, male). Data are taken from the study cohort in  \cite{King:isoprene} and \cite{King:2009}. Exercise segments are shaded in grey.
 (First column): Two legged ergometer experiment.
Protocol: 5~min resting - 15~min exercise (75~W) - 3~min resting - 15~min exercise (75~W) - 12~min resting - 5~min exercise (75~W) - 5~min resting. (Second column): One legged ergometer experiment. Protocol: 3~min resting - 10~min left leg (50~W) - 4~min resting - 2~min left leg (50~W) - 4~min resting - 7~min right leg (50~W). (Third column): Arm/two-legged ergometer experiment. Protocol: 2~min resting - 5~min arm-crank training (2.5~kg/arm) - 4~min resting - 2~min arm-crank training (2.5~kg/arm) - 4~min resting - 4~min two-legged exercise (60~W). }
 \label{fig:exercise}
\end{figure}

The experimental data presented in the following are taken from the study cohort in  \cite{King:isoprene} and \cite{King:2009}, where five (age 27-34 years, 4 male, 1 female) and eight (age 25-30, 5 male, 3 female) normal healthy volunteers have been investigated, respectively. Figure \ref{fig:exercise} shows  representative experimental results for one single volunteer (27 years, male).
 Breath isoprene concentrations are assessed by means of a real-time setup designed for synchronized measurements of exhaled breath VOCs (using PTR--MS) as well as a variety of respiratory and hemodynamic parameters. For a detailed description of the instrumental setup  the interested reader is referred to \cite{King:2009}.
All results were obtained in conformity with the Declaration of Helsinki and with approval by the Ethics Commission of Innsbruck Medical University.

Due to its volatility and low affinity for blood (as reflected by a small blood:gas partition coefficient  $\hen=$ 0.75 at body temperature  \cite{Filser:1996}), isoprene exchange occurs mainly in the alveolar region.  Physical activity causes marked changes in exhaled breath isoprene concentrations of humans   ~\cite{King:2009,Sent:2000,Karl:2001}. Real-time measurements during moderate workload ergometer challenges were provided by Karl et al.~\cite{Karl:2001} and King et al.~\cite{King:2009}, showing an initial increase within the first minute of exercise by a factor of $\sim 2-3$  in mixed exhaled breath  \cite{Karl:2001}, and by a factor of $\sim 3-4$  in end-tidal breath \cite{King:2009}, respectively.
Physiologically, this effect could be attributed to 
 functional changes in the lung such as redistribution of ventilation and/or perfusion, recruitment and distension of pulmonary capillaries or changes in mixed venous concentration due to depletion  of an isoprene buffer tissue.

In normal subjects, the more uniform topographical distribution of pulmonary blood flow during exercise results in a relatively uniform matching of ventilation and perfusion throughout the lung ~\cite{West:Book,Levitzky:Book}.
Since in moderate to severe exercise ventilation increases more than perfusion, the overall ventilation-perfusion ratio increases from a value of about 1 at rest to a range of 2 to 4 \cite{Levitzky:Book}. Other factors being equal, this effect would cause a decrease in the exhaled breath concentrations of low soluble VOCs according to  Farhi's equation~(\ref{farhi}). Interestingly, exhaled isoprene profiles during workload challenges drastically depart from this predicted trend, see~Figure \ref{fig:exercise}.
In contrast, exhaled breath profiles of butane (also a blood-borne endogenous VOC) under dynamical exercise obey  the qualitative behavior anticipated from Farhi's equation (\ref{farhi}) \cite{King:PhysiolMeas}, even though both compounds are comparable in terms of their physico-chemical properties. In particular, both butane and isoprene have similar molecular weights and similar blood:gas partition coefficients, and therefore the diffusing capacity of the lung for both  gases should be similar according to \textit{Graham's law} (see \ref{sec:diff}). This discrepancy suggests that some compound-specific (release) mechanism needs to be taken into account for clarifying the physiological flow of isoprene. 
 
The first column of Figure \ref{fig:exercise} presents measurement results during a 2-legged ergometer exercise with three subsequent workload phases at 75 W, and pauses of 3 min and 12 min, respectively. 
The breath isoprene concentration profiles during these three workload phases display similar shape. Nevertheless, the height of the characteristic exercise peak is significantly higher during the first phase of workload, despite an almost identical behavior of cardiac output and alveolar ventilation throughout all exercise segments. This indicates the existence of a buffer tissue compartment for isoprene. The concentration peak observed during the first workload phase is restored if the pauses between workload phases are extended: Approximately one hour of rest is necessary to achieve a complete recovery of the initial peak height \cite{King:2009}.

The second column of Figure \ref{fig:exercise} shows representative results in response to one-legged ergometer exercise at 50~W.  After 10 minutes of pedaling with the left leg, breath isoprene profiles closely  resemble the two-legged case and a clear washout effect emerges, yielding a lower peak height when continuing the exercise with the same leg. However, if the working limb is switched to the right leg after a short break, an immediate recovery of the initial peak size can be observed. 
The observation that the rise in cardiac output and alveolar ventilation is of comparable order in all three phases of exercise appears to exclude functional changes of the lung (redistribution of ventilation-perfusion ratio, distension of pulmonary capillaries) as the main cause for the peak-shaped isoprene profile at the onset of exercise and strongly supports the hypothesis of a peripheral source affecting breath isoprene output.

The experiment presented in the third column of Figure \ref{fig:exercise} allows for a more precise specification of the location of the above-mentioned tentative isoprene buffer, namely skeletal muscles, by stimulating single muscle groups with distinct masses.  
While repeated dynamic arm-crank exercises with an intermediate pause of 4~minutes yield the same washout effect as described above, subsequent two-legged  ergometer exercise at 60~W yields a much higher peak compared to arm-crank exercise, despite a similar behavior of the ventilation-perfusion ratio within all three workload segments. 
We attribute this effect to  the smaller mass of arm musculature as well as to the smaller fractional perfusion of the arm muscle group as compared to leg musculature. This experiment proposes that a major part of isoprene variability during exercise conditions can be ascribed to an increased fractional perfusion of the working locomotor muscles, eventually leading to higher isoprene concentrations in mixed venous blood at the onset of physical activity. 

This rationale is in accordance with the predominant physiological role of  working muscle during exercise.   Collectively, the skeletal muscles constitute up to 40-45\% of body weight, which is more than any other single organ. At rest, about 10--15\% of cardiac output is distributed to skeletal muscle, while during strenuous exercise skeletal muscle may receive more than 80\% of total blood flow, thus rendering it as one of the major factors in overall cardiovascular hemodynamics \cite{Mohrman:book}.

\section{The model}
\begin{figure}[H]

\centering
\begin{tabular}{c}
\includegraphics[scale=1]{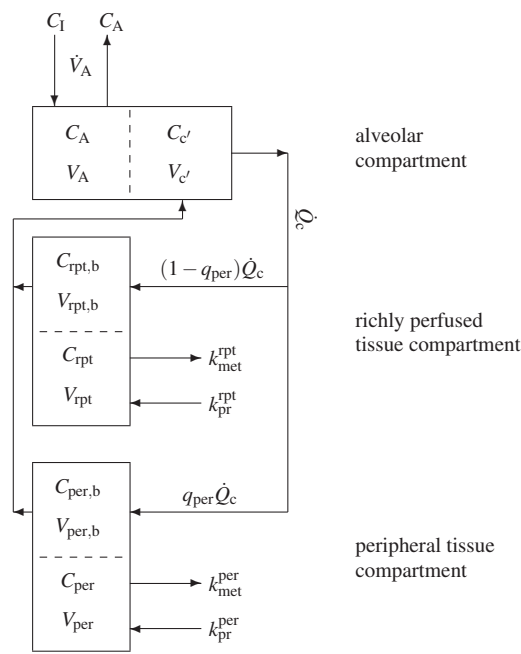}
\end{tabular}
\caption{Sketch of the model structure. The body is divided into three distinct functional units: alveolar/end-capillary compartment (gas exchange), richly perfused tissue (metabolism and production) and peripheral tissue (storage, metabolism and production). Dashed boundaries indicate a diffusion equilibrium. Abbreviations connote as in Table~\ref{table:param}.}\label{fig:model_struct}
\end{figure}

In light of the experiments outlined above
the mathematical model of isoprene distribution presented here is based on the assumption of a peripheral source of isoprene in the body. 
The lung is assumed to be a single homogenous compartment where an instantaneous equilibration between the arterial concentration leaving the lungs (corresponding to the end-capillary blood concentration $C_{\mathrm{c'}}$) is reached. This is considered to be a reasonable assumption for inert gases as mentioned in Section \ref{sec:comp}. 
The body is  subdivided into two homogenous functional units: a richly perfused tissue compartment  (including intestines, liver, brain, connective muscles and skin) and a peripheral tissue compartment. The model structure is sketched in Figure \ref{fig:model_struct}.  While  production and metabolic elimination of isoprene occurs in both compartments, the peripheral tissue compartment takes the role of an  isoprene buffer and is postulated to contain the working muscle compartment, which receives  a disproportionally high fraction of the systemic blood flow as soon as exercise starts.
During rest, the peripheral compartment is characterized by high isoprene concentrations resulting from extrahepatic production with a constant rate. However, due to the minute fractional blood flow to these tissues, mixed venous concentrations are mainly governed by the lower isoprene content in venous blood returning from the richly perfused tissue group. At the start of exercise, the fractional perfusion in the periphery increases and the mixed venous concentrations become dominated by peripheral venous return. The isoprene concentration peak visible in breath hence is  considered as a consequence of the corresponding increase in the underlying mixed venous concentration. The model equations are given in \ref{sec:peripheral},  while a detailed description of the model structure, its validation and estimation can be found in \cite{King:isoprene}.

\section{Discussion}

The interplay between mathematical modeling and theory-driven
experimentation has finally led us to a mathematical description of
isoprene kinetics which respects all current data and which furthermore
allows for new predictions that can be tested by adequate experiments.

The origin of endogenous isoprene in humans is mainly attributed to the mevalonate pathway of cholesterol biosynthesis \cite{Karl:2001,Deneris:1985,Stone:1993}. The synthesis of mevalonate is inhibited by statins, which are used in the treatment and prevention of cardiovascular diseases by lowering serum cholesterol levels. The primary mechanism of action exerted by statin therapy is to competitively inhibit the activity of HMG-CoA reductase, which is the rate limiting enzyme in cholesterol synthesis \cite{Farmer:2003}. Correspondingly, it has been reported that the administration of specific statins (such as lovastatin and atorvastatin) causes a proportional decline in breath isoprene concentrations and serum cholesterol levels  \cite{Karl:2001,Stone:1993}. On the other hand, several studies have pointed out myotoxic effects to be a major adverse reaction of statin therapy, leading to the inhibition of skeletal muscle activity  \cite{Farmer:2003,Evans:2002, Walravens:1989}.

The extent to which the mevalonate pathway accounts for isoprene formation under physiological conditions is still a matter of debate \cite{Miekisch:2004, Taucher:1997}. In particular, the above-mentioned pathway rests on the acid-catalyzed solvolysis of dimethylallyl-diphosphate in the liver which may be insignificant at physiological pH \cite{Silver:1991,Sharkey:1996}. 
In many plants, the formation of isoprene is enzyme-catalyzed, and isoprene seems to play a role in the heat protection of leafs \cite{Silver:1995}. Moreover, some indications of an isoprene synthase in bovine liver have been put forward in \cite{Sharkey:1996}. Hence, it cannot be excluded that an isoprene synthase is present in human tissues. However that may be, we suggest an additional peripheral origin of isoprene through the same basic biochemical pathway. 
Specifically, by adopting the assumption of a muscular source of isoprene, statin-induced lowering of isoprene concentration in the muscles may turn out as the main cause for the reduction of endogenous isoprene formation and the subsequent decrease of breath isoprene concentrations. 
An analogous argument might also give further insights into the surprising lack of statistically significant correlations between breath isoprene levels and blood cholesterol levels \cite{Kushch:2008, Sharkey:1996}.

Similarly, the proposed mechanisms may explain the age dependency of breath isoprene levels by correlating the latter to changes in the individual muscle mass. 
The characteristic median breath isoprene concentration in adults under resting conditions is about 100 ppb \cite{Kushch:2008}, whereas young children have demonstrably lower isoprene levels \cite{Taucher:1997,Lindinger:1998-1,Smith:2010}.
Breath isoprene concentrations have been  reported to be non-detectable in the breath of neonates, while steadily increasing in teenagers and reaching a plateau level in the middle age \cite{Smith:2010,Book:MY-Risby}. Moreover, isoprene levels in older people appear to decrease \cite{Kushch:2008,Book:MY-Risby}.

While the cause and effect relationships proposed above remain speculative, the modeling study presented in  \cite{King:isoprene} and this paper  certainly  yields new interesting perspectives on previous experimental findings. 
Several experiments have been summarized here demonstrating that the peak-shaped behavior of end-tidal isoprene during exercise is most probably associated with the increased activity and perfusion of the skeletal muscle group. Former investigations with respect to the age dependency of isoprene output and its reduction under statin therapy nicely fit into this rationale. Moreover, the present work suggests that the major part of physiologically formed isoprene stems from the skeletal musculature rather than hepatic production.

\section{Acknowledgements}
The research leading to these results has received funding from the European CommunityÕs Seventh Framework Programme (FP7/2007-13) under grant agreement No.~217967. We greatly appreciate funding from the Austrian Federal Ministry for Transport, Innovation and Technology (BMVIT/BMWA, project 818803, KIRAS). Julian King is a recipient of a DOC fellowship of the Austrian Academy of Sciences at the Breath Research Institute. Gerald Teschl and Julian King acknowledge support from the Austrian Science Fund (FWF) under Grant No.~Y330. We greatly appreciate the generous support of Vorarlberg and its governor Dr. Herbert Sausgruber.
 
\appendix
\section{}
\label{sec:appendix}

The following paragraphs briefly summarize the fundamental physical principles underlying the present model derivation. Further details can be found in \cite{King:isoprene}.

\subsection{Mass transport in the lung}\label{sec:general}
In the basic model describing the pulmonary gas exchange of blood borne-inert gases, the lung is considered to be a homogenous single compartment with a fixed storage volume $V_\mathrm{A}$. An instantaneous equilibrium between end-capillary  blood (corresponding to arterial blood) and alveolar air is assumed, leading to a directly proportional relationship (Henry's law) between the arterial blood concentration $\cart$ and the alveolar air concentration $\calv$, viz.,  
\begin{equation}\label{eq:cart}
\cart =  \hen\, \calv.
\end{equation}
Here, $\hen$ is the dimensionless substance specific partition coefficient between blood and air.
In view of this diffusion equilibrium, the alveolar compartment capacity is governed by the effective storage volume $\valv:=V_{\mathrm{A}}+ \hen \, V_{\mathrm{c'}} $, where $V_{\mathrm{c'}}$ represents the capillary blood volume.

The rate of accumulation of a gas within this volume equals the difference between supply and elimination due to perfusion $\qc$ and ventilation $\qalv$, where both quantities are regarded as continuous parameters \footnote{i.e., only   averages over time periods are considered}. This relationship is described by the following mass balance equation (see, e.g., \cite{Susanne})

\begin{equation}\label{eq:lung}
\valv \frac{d}{dt} \calv = \qc \left(\cven - \cart   \right) + \qalv \left(\cinh - \calv \right),
\end{equation}
where $\cven$ and $\cinh$  are the concentrations in mixed venous blood and inhaled air, respectively. The expired air is assumed to have the same concentration as the alveolar air.

\subsection{A three compartment model for isoprene exchange}\label{sec:peripheral}
In the gas exchange model for isoprene adopted here, the mass balance equation for the lung  corresponds to Equation (\ref{eq:lung}), adding the assumption that inhaled air only contains negligible amounts of isoprene, i.e., $\cinh=0$.

The systemic part consists of  two well mixed 
compartments: a richly perfused tissue compartment (including intestines, brain, connective muscles, skin)  and a peripheral tissue compartment. Both compartments are active in the sense that production as well as metabolic elimination of isoprene occurs. The venous blood concentration leaving the compartment is considered to be in equilibrium with the respective tissue concentration at every instant t (venous equilibrium). Based on this assumption, the storage capacities in richly perfused and peripheral tissue can again be expressed as effective volumes $\vrpt:=V_{\mathrm{rpt}}+ \lrpt\, V_{\mathrm{rpt,b}} $ and  $\vper:=V_{\mathrm{per}}+ \lper \, V_{\mathrm{per,b}} $, respectively. Here, $V_{\mathrm{\star}}$ and $V_{\mathrm{\star,b}}$ denote the volumes of intracellular space and vascular blood, respectively, while $\lambda_{\mathrm{b:\star}}$   is the blood:tissue partition coefficient of the corresponding compartment \footnote{here $\star$ denotes $\mathrm{rpt}$ or $\mathrm{per}$}. 
Mass balance equations read

\begin{equation}\label{eq:rpt}
\vrpt\frac{\di\crpt}{\di t}=(1-\qper)\qc(\cart -\lrpt\crpt)+\prl-\ml\lrpt\crpt\,,
\end{equation} 
for richly perfused tissue and
\begin{equation}\label{eq:per}
\vper\frac{\di\cper}{\di t}=\qper\qc(\cart -\lper\cper)+\prm-\mm\lper\cper\,,
\end{equation}
for peripheral tissue, with the kinetic rate constants ${k_{\mathrm{pr}}^\mathrm{\star}}$ and  ${k_{\mathrm{met}}^\mathrm{\star}}$ describing production and metabolic elimination, respectively, and $q_{\mathrm{\star}}$ denoting a fractional blood flow.

In order to capture the redistribution of the systemic perfusion during ergometer exercise, the fractional blood flow $\qper \in (0,1)$ to peripheral tissue is assumed to resemble the fractional blood flow to both legs, which is postulated to increase with cardiac output according to
\begin{equation}\label{eq:qper}\qper(\qc):=\qper^{\mathrm{rest}}+(\qper^{\mathrm{max}}-\qper^{\mathrm{rest}})\times \\ \big(1-\exp{(-\tau\,\max\{0,\frac{\qc-\qc^{\mathrm{rest}}}{\qc^{\mathrm{rest}}}\})}\big), \end{equation}
where $\tau>0$ is a constant.
The associated concentrations in mixed venous and arterial blood are  given by 
\begin{equation}\label{eq:cven}\cven:=(1-\qper)\lrpt\crpt+\qper\lper\cper\end{equation}
and Equation~(\ref{eq:cart}), respectively.
Moreover, we state that the measured (end-tidal) isoprene concentration equals the alveolar level, i.e.,
\begin{equation}\label{eq:meas}\cmeas=\calv.\end{equation}

\subsection{Diffusion}\label{sec:diff}

The diffusive flux $J$ of gases through a barrier separating two regions characterized by distinct partial 
pressures $P_j,\, j=1,2$ is governed by \textit{Fick's first law} of diffusion
\begin{equation}\label{Ficks law}
J = \frac{\solubility \diffusivity \area}{\ell} \left(P_{1}-P_{2}\right).
\end{equation}
While moving between the alveolar space and capillary blood, each gas is subject to the same anatomically related limitations (cross sectional area  $\area$ and thickness of the membrane  $\thickness$). Each gas, however, has a different solubility  $\solubility$  and diffusivity  $\diffusivity$  in the membrane barrier.

Generally, all anatomically and gas related parameters in Equation  (\ref{Ficks law}) are lumped together under the term \textit{diffusing capacity of the lung} ($\diffcap$). At the alveolar-capillary interface Equation (\ref{Ficks law}) hence simplifies to

\begin{equation}\label{simplified Ficks law}
J=\diffcap (\palv-\pblood)\,,
\end{equation}
where $\palv$ and $\pblood$ are the partial pressures in alveolar air and capillary blood, respectively. 

\textit{Graham's law} states that the diffusivity of a gas is inversely proportional  to the square root of its molecular weight  $\mathrm{MW}$. Therefore, the relative diffusing capacities of the lung for two distinct gases can be estimated via

\begin{equation}\label{relative diffcap}
\frac{D_{gas,1}}{D_{gas,2}}=\frac{\beta_{gas,1}}{\beta_{gas,2}} \times \frac{\sqrt{\mathrm{MW}_{gas,2}}}{\sqrt{\mathrm{MW}_{gas,1}}}\,.
\end{equation}
\section*{Nomenclature}
\begin{table}[H]
\centering 
\footnotesize
\begin{tabular}{|lc|}\hline
 {\large\strut}\textbf{Parameter}&\textbf{Symbol} \\ \hline \hline
{\large\strut}\textit{Concentrations} &   \\ 
{\large\strut}  alveoli &$\calv$  \\
{\large\strut}  end-capillary &$C_{\mathrm{c'}}$   \\
{\large\strut}  arterial &$\cart$  \\
{\large\strut}  mixed-venous &$\cven$  \\
{\large\strut}  richly perfused tissue (rpt) &$\crpt$   \\
{\large\strut}  peripheral tissue &$\cper$   \\
{\large\strut}  inhaled (ambient) &$\cinh$   \\
{\large\strut}\textit{Compartment volumes} &   \\
{\large\strut}  alveoli &$V_{\mathrm{A}}$  \\
{\large\strut}  end-capillary &$V_{\mathrm{c'}}$  \\
{\large\strut}  rpt &$V_{\mathrm{rpt}}$  \\
{\large\strut}  blood rpt &$V_{\mathrm{rpt,b}}$  \\
{\large\strut}  peripheral tissue &$V_{\mathrm{per}}$   \\
{\large\strut}  blood peripheral tissue & $V_{\mathrm{per,b}}$   \\
{\large\strut}  ambient &$\vinh$   \\
{\large\strut}\textit{Fractional blood flows} &   \\
{\large\strut}  periphery (both legs)&$\qper$   \\
{\large\strut}  maximal &$\qper^\mathrm{max}$  \\
{\large\strut}  nominal (rest) &$\qper^\mathrm{rest}$  \\
{\large\strut}  constant in Eq.~\ref{eq:qper} &$\quad \tau$   \\
{\large\strut}\textit{Partition coefficients} &   \\
{\large\strut}  blood:air &$\hen$  \\
{\large\strut}  blood:rpt &$\lrpt$   \\
{\large\strut}  blood:peripheral tissue &$\lper$  \\
{\large\strut}\textit{Rate constants} &   \\
{\large\strut}  hepatic metabolic rate &$\ml$   \\
{\large\strut}  extrahepatic metabolic rate &$\mm$   \\
{\large\strut}  production rpt &$\prl$   \\
{\large\strut}  production peripheral tissue &$\prm$   \\
\hline
\end{tabular}
\caption{List of symbols }\label{table:param}
\end{table}

\section*{References}
\bibliographystyle{unsrt}

\bibliography{BGABib}

\end{document}